# On the distance and algorithms of strong product digraphs *

Haoran Yin[1,2,3]    Feng Li[1,†]

1 College of Computer Science, Qinghai Normal University, Xi'ning, 810008, P.R.China

2 Tibetan Information Processing and Machine Translation Key Laboratory of Qinghai Province, Xi'ning, 810008, P.R.China

3 Key Laboratory of Tibetan information Processing, Ministry of Education, Xi'ning, 810008, P.R.China

**Abstract**

Strong product is an efficient way to construct a larger digraph through some specific small digraphs. The large digraph constructed by the strong product method contains the factor digraphs as its subgraphs, and can retain some good properties of the factor digraphs. The distance of digraphs is one of the most basic structural parameters in graph theory, and it plays an important role in analyzing the effectiveness of interconnection networks. In particular, it provides a basis for measuring the transmission delay of networks. When the topological structure of an interconnection network is represented by a digraph, the average distance of the directed graph is a good measure of the communication performance of the network. In this paper, we mainly investigate the distance and average distance of strong product digraphs, and give a formula for the distance of strong product digraphs and an algorithm for solving the average distance of strong product digraphs.

**Keywords**: Digraphs; Strong product; Distance; Average distance; Algorithm

## 1  Introduction

In this paper, we only consider finite and simple directed graphs (or digraphs). L-

* Project supported by the National Natural Science Foundation of China (No. 11551002), The Natural Science Foundation of Qinghai Province (2019-ZJ-7093).
† Corresponding author. E-mail address: li2006369@126.com; yinhr163@163.com.

et $G = (V(G), E(G))$ be a digraph with the vertex set $V(G)$ and the edge set $E(G)$. A directed walk in $G$ is a finite and non-empty sequence composed of alternating vertices and edges:

$$W = v_0 e_1 v_1 e_2 v_2 \cdots e_k v_k,$$

such that for $1 \leq i \leq k$, the tail of edge $e_i$ is $v_{i-1}$ and the head is $v_i$. $W$ is called a directed walk from vertex $v_0$ to $v_k$, denoted by $(v_0, v_k)$ walk. $k$ is called the length of the directed walk $W$, denoted by $\varepsilon(W) = k$. If all the vertices $v_0, v_1, \cdots, v_k$ in the walk $W$ are different from each other, then $W$ is called a path. Since the simple digraph has at most one edge from one vertex to the other, the walk $W$ is determined by its vertex sequence $v_0, v_1, \cdots, v_k$. Therefore, the walk of simple digraphs can simply use its vertices sequence to represent. For notations and terminologies not defined here we follow Ref. [1].

Let $G$ be a digraph and $x, y \in V(G)$. The length of the shortest $(x, y)$ path in $G$ is called the distance from $x$ to $y$, denoted by $d_G(x, y)$. The maximum distance between any two vertices in $G$ is called the diameter of $G$, denoted by $diam(G)$. The transmission delay from the source to the destination in interconnection networks is a useful parameter to measure the performance of networks. If a digraph represents the topological structure of a store-and-forward interconnection network, the distance and diameter of the digraph provide a basis for measuring the network's transmission delay [2]. Therefore, many authors have studied the distance and diameter of graphs, and readers can refer to the recent literatures [3-6].

Although the diameter of digraphs can effectively measure the transmission delay of networks, the diameter is the maximum distance between any two vertices in digraphs, that is, the diameter of the digraphs is only a measure of the worst case of network communication performance. However, in practical applications, there are always very few vertex pairs with large distance in networks. Therefore, the average distance is better than the diameter to measure the overall communication performance of networks. Let $G$ be a strongly connected digraph of order $n$, where $n \geq 2$. The average distance of $G$, denoted by $\mu(G)$, is defined as

$$\mu(G) = \frac{1}{n(n-1)} \sum_{x,y \in V(G)} d_G(x, y).$$

It can be easily seen that $\mu(G) \geq 1$, because $\mu(G)$ is the arithmetic average of all non-zero distances. $\mu(G) = 1$ if and only if $G$ is a complete digraph. Since the average distance of graphs is important in the field of graph theory and network an-alysis, many researchers have studied it. For results of the study on the average distance of general graphs, readers can refer to Plesnik [7]. Most of these results are upper and lower bounds of the average distance represented by some graph theory parameters. The average distance of some specific graphs has also been studied by many researchers. For example, Klavžar [8] studied the average distance and other structural properties of exchanged hypercubes. For more results on the average distance of graphs, readers can read papers [9-13].

The concept of strong product was first introduced by sabidussi [14], and he also introduced the concept of Cartesian product and weak product. The product of graphs is not only an important operation for graphs, but also an important and effective method for designing large-scale interconnection networks. In this paper, we consider the strong product of digraphs. Let $G_1 = (V(G_1), E(G_1))$ and $G_2 = (V(G_2), E(G_2))$ be two directed graphs. The strong product of $G_1$ and $G_2$, denoted by $G_1 \boxtimes G_2$, is a directed graph with a vertex set $V(G_1 \boxtimes G_2) = V(G_1) \times V(G_2)$. For any two distinct vertices $x_1 x_2, y_1 y_2 \in V(G_1) \times V(G_2)$, there is a directed edge from $x_1 x_2$ to $y_1 y_2$ if and only if either $x_1 = y_1$ and $(x_2, y_2) \in E(G_2)$, or $x_2 = y_2$ and $(x_1, y_1) \in E(G_1)$, or $(x_1, y_1) \in E(G_1)$ and $(x_2, y_2) \in E(G_2)$. $G_1$ and $G_2$ are called the factor digraphs of $G_1 \boxtimes G_2$. We also call the digraphs constructed by strong product as strong product digraphs. For further information and results about the strong product, readers can refer to articles [15-19].

The large digraph constructed by the strong product method contains the factor digraphs as its subgraphs, and retains many good properties of the factor digraphs, such as regularity, connectivity, embeddability, etc. (see Ref. [20, 21]). Since a strong product digraph is constructed by some specific digraphs, the topological structure of the factor digraphs determines the topological structure of the strong product digraph. We mainly study the distance and average distance of strong product digraphs, and give a formula for the distance of strong product digraphs and an algorithm for solving the average distance of strong product digraphs.

## 2 The distance of strong product digraphs

Determining the distance between any two vertices in strong product digraphs has important significance in graph theory and network analysis, and can also provide theoretical basis for studying the diameter, radius and Wiener index of strong product digraphs. According to the definition of strong product, we can prove the following theorem by using the construction method.

**Theorem 2.1** Let $G_1$ and $G_2$ be two strongly connected digraphs. Let $x = x_1 x_2$ and $y = y_1 y_2$ be two vertices of the strong product digraph $G_1 \boxtimes G_2$, where $x_1, y_1 \in V(G_1)$, $x_2, y_2 \in V(G_2)$. Then
$$d_{G_1 \boxtimes G_2}(x, y) = max\{ d_{G_1}(x_1, y_1), d_{G_2}(x_2, y_2) \}.$$

Proof. The theorem is clearly true when $x_1 = y_1$ or $x_2 = y_2$. So next we assume $x_1 \neq y_1$ and $x_2 \neq y_2$. Let $d_{G_1}(x_1, y_1)$ be the distance from vertex $x_1$ to vertex $y_1$ in the digraph $G_1$. Since $G_1$ is strongly connected, there must be a shortest directed path $(x_1, y_1)$ in $G_1$, denoted by $P$.
$$P: x_1 = g_1, g_2, \cdots, g_m, \cdots, g_n = y_1.$$
Let $d_{G_2}(x_2, y_2)$ be the distance from vertex $x_2$ to vertex $y_2$ in the digraph $G_2$. Since $G_2$ is strongly connected, there must be a shortest directed path $(x_2, y_2)$ in $G_2$, denoted by $Q$.
$$Q: x_2 = t_1, t_2, \cdots, t_r, \cdots, t_s = y_2.$$
Next, we distinguish the following cases:

Case (1): If $n = s$, that is, the lengths of the two shortest directed paths $P$ and $Q$ are equal, then by the definition of strong product, there is a directed $(x_1 x_2, y_1 y_2)$ path in the digraph $G_1 \boxtimes G_2$ as follows:
$$W: x_1 x_2 = g_1 t_1, g_2 t_2, \cdots, g_n t_s = y_1 y_2.$$
The length of $W$ is equal to $P$ and $Q$, i.e. $\varepsilon(W) = \varepsilon(P) = \varepsilon(Q)$. We assert that $W$ is a shortest directed path from $x_1 x_2$ to $y_1 y_2$. Otherwise, let
$$L: x_1 x_2, u_1 u_2, \cdots, v_1 v_2, y_1 y_2.$$
be the shortest directed path from $x_1 x_2$ to $y_1 y_2$ in $G_1 \boxtimes G_2$. We take the second coordinates of all the vertices in $L$ in the original order and remove all the adjacent repeated vertices. Thus we get a $(x_2, y_2)$ walk in the digraph $G_2$, denoted by $Q'$. Then
$$\varepsilon(W) > \varepsilon(L) \geq \varepsilon(Q') \geq \varepsilon(Q) = \varepsilon(W).$$
Contradiction. Therefore, $W$ is the shortest directed path from vertex $x_1 x_2$ to $y_1 y_2$ in the strong product digraph $G_1 \boxtimes G_2$. Then we have
$$d_{G_1 \times G_2}(x, y) = d_{G_1}(x_1, y_1) = d_{G_2}(x_2, y_2).$$

Case (2): If $n > s$, that is, the length of $P$ is longer than the length of $Q$, then by the definition of strong product, there is a directed path $(x_1x_2, y_1y_2)$ in the digraph $G_1 \boxtimes G_2$ as follows:

$$R: x_1x_2 = g_1t_1, g_2t_2, \cdots, g_st_s, g_{s+1}t_s, \cdots, g_nt_s = y_1y_2.$$

The length of $W$ is equal to $P$, i.e. $\varepsilon(R) = \varepsilon(P)$. We assert that $R$ is a shortest path from $x_1x_2$ to $y_1y_2$. Otherwise, let

$$T: x_1x_2, a_1a_2, \cdots, b_1b_2, y_1y_2.$$

be the shortest directed path from $x_1x_2$ to $y_1y_2$ in $G_1 \boxtimes G_2$. We take the first coordinates of all the vertices in $T$ in the original order and remove all the adjacent repeated vertices. Thus we get a $(x_1, y_1)$ walk in the digraph $G_1$, denoted by $P'$. Then

$$\varepsilon(R) > \varepsilon(T) \geq \varepsilon(P') \geq \varepsilon(P) = \varepsilon(R).$$

Contradiction. Therefore, $R$ is the shortest directed path from vertex $x_1x_2$ to $y_1y_2$ in the strong product digraph $G_1 \boxtimes G_2$. Then we have

$$d_{G_1 \times G_2}(x, y) = d_{G_1}(x_1, y_1).$$

Case (3): When $n < s$, by using a method similar to that of Case(2), we can get the distance from $x_1x_2$ to $y_1y_2$ is:

$$d_{G_1 \times G_2}(x, y) = d_{G_2}(x_2, y_2).$$

Combining the above cases, the proof of the theorem is complete.

For the strong product digraph of $n$ digraphs $G_1, G_2, \cdots, G_n$, by using a method similar to Theorem 2.1, the vertex distance formula of $G_1 \boxtimes G_2 \boxtimes \cdots \boxtimes G_n$ can also be easy to get. Thus we have the following theorem:

**Theorem 2.2** Let $G_1, G_2, \cdots, G_n$ be strongly connected digraphs. Let $x = x_1, x_2, \cdots, x_n$ and $y = y_1y_2, \cdots, y_n$ be two vertices of the strong product digraph $G_1 \boxtimes G_2 \boxtimes \cdots \boxtimes G_n$, where $x_i, y_i \in V(G_i)$, $i = 1, 2, \cdots, n$. Then

$$d_{G_1 \boxtimes G_2 \boxtimes \cdots \boxtimes G_n}(x, y) = max\{ d_{G_1}(x_1, y_1), d_{G_2}(x_2, y_2), \cdots, d_{G_n}(x_n, y_n) \}.$$

## 3  The algorithm to solve average distance

We will start with the concept of adjacency matrix. Let $G$ be a digraph with a vertex set $V(G) = \{x_1, x_2, \cdots, x_n\}$. The adjacency matrix of $G$ is a $n \times n$ matrix, de-noted by $A(G) = (a_{ij})$, where
$$a_{ij} = \begin{cases} 1, & if\ (x_i, y_j) \in E(G), \\ 0, & if\ (x_i, y_j) \notin E(G). \end{cases}$$

Obviously, the adjacency matrix of a simple digraph is a $(0,1)$ matrix in which the diagonal elements are all zero. The adjacency matrix of a digraph is another representation of the digraph, which is stored in computers in this form.

Let $G_1$ and $G_2$ be two strongly connected digraphs, and $x_1x_2$ and $y_1y_2$ be any two vertices of the strong product digraph $G_1 \boxtimes G_2$, where $x_1, y_1 \in V(G_1)$ and $x_2, y_2 \in V(G_2)$. Let $V_i = V(G_i)$, $v_i = |V(G_i)|$ where $i = 1,2$. Let $V = V_1 \times V_2$. Then the average distance of the strong product digraph $G_1 \boxtimes G_2$ can be expressed as:

$$\mu(G_1 \times G_2) = \frac{1}{v_1 v_2 (v_1 v_2 - 1)} \sum_{x_1 x_2, y_1 y_2 \in V} d_{G_1 \times G_2}(x_1 x_2, y_1 y_2).$$

To calculate the average distance of the strong product digraph $G_1 \boxtimes G_2$, we must first calculate the distance between all pairs of vertices in $G_1 \boxtimes G_2$. For convenience, let

$$\sigma(G_1 \times G_2) = \sum_{x_1 x_2, y_1 y_2 \in V} d_{G_1 \times G_2}(x_1 x_2, y_1 y_2).$$

Then we have

$$\begin{aligned}\sigma(G_1 \times G_2) &= \sum_{x_1 x_2, y_1 y_2 \in V} d_{G_1 \times G_2}(x_1 x_2, y_1 y_2) \\ &= \sum_{x_1 x_2 \in V} \sum_{y_1 y_2 \in V} d_{G_1 \times G_2}(x_1 x_2, y_1 y_2) \\ &= \sum_{x_1 \in V_1} \sum_{y_1 \in V_1} \sum_{x_2 \in V_2} \sum_{y_2 \in V_2} d_{G_1 \times G_2}(x_1 x_2, y_1 y_2).\end{aligned}$$

According to the distance formula of strong product digraphs:

$$d_{G_1 \times G_2}(x_1 x_2, y_1 y_2) = max\{ d_{G_1}(x_1, y_1), d_{G_2}(x_2, y_2) \}.$$

We can know that in order to calculate the average distance of the strong product digraph $G_1 \boxtimes G_2$, we need to compare the distance between any two vertices in $G_1$ and the distance between any two vertices in $G_2$ (total $v_1^2 v_2^2$ times). This means it is very difficult to derive a formula for the average distance of strong product digraphs. Therefore, we have designed an algorithm that calculates the average distance of the strong product digraph $G_1 \boxtimes G_2$ in polynomial time.

**Algorithm**: The average distance of strong product digraphs

**Data**: Adjacency matrix $A_1$ and number of vertices $v_1$ of digraph $G_1$. Adjacency matrix $A_2$ and number of vertices $v_2$ of digraph $G_2$.

**Result**: The average distance $\mu(G_1 \boxtimes G_2)$ of the strong product digraph $G_1 \boxtimes G_2$.

**Program process**

  **S1** begin main program
  **S2** call averageDistance($A_1, v_1, A_2, v_2$)
  **S3** end main program
  **S4** procedure averageDistance($A_1, v_1, A_2, v_2$)
  **S5** begin
    (1) $D_1 := A_1$ and $D_2 := A_2$
    (2) call distanceMatrix($D_1, v_1$)
    (3) call distanceMatrix($D_2, v_2$)
    (4) $\sigma(G_1 \boxtimes G_2) := 0$
    (5) for $x := 1$ to $v_1$
        for $y := 1$ to $v_1$
           for $m := 1$ to $v_2$
               for $n := 1$ to $v_2$
                  (5.1) if $D_1[x, y] > D_2[m, n]$
                  (5.2) then $\sigma(G_1 \boxtimes G_2) := \sigma(G_1 \boxtimes G_2) + D_1[x, y]$
                  (5.3) else $\sigma(G_1 \boxtimes G_2) := \sigma(G_1 \boxtimes G_2) + D_2[m, n]$
                  (5.4) end if
               end for
           end for
        end for
      end for

    (6) $\mu(G_1 \boxtimes G_2) := \sigma(G_1 \boxtimes G_2)/(v_1 v_2 (v_1 v_2 - 1))$

  **S6** end
  **S7** procedure distanceMatrix($D, v$)
  **S8** begin
    (1) foreach off-diagonal element $e$ in $D$ do
        (1.1) if $e = 0$
        (1.2) then $e := \infty$
        (1.3) end if
    (2) end for
    (3) for $k := 1$ to $v$
        for $m := 1$ to $v$
           for $n := 1$ to $v$
               (3.1) if $D[m, n] > D[m, k] + D[k, n]$
               (3.2) then $D[m, n] := D[m, k] + D[k, n]$
               (3.3) end if
           end for
        end for

end for
    **S9** end

---

In the above algorithm, $\sigma(G_1 \boxtimes G_2)$ represents the sum of the distances between all ordered pairs of vertices of the strong product digraph $G_1 \boxtimes G_2$. For $i = 1, 2$, $D_i$ is a matrix formed by the distances between all ordered pairs of vertices of in $G_i$ as its elements. That is, the value of the element $D_i[x][y]$ is the distance from the vertex $x$ to the vertex $y$ in $G_i$.

The steps of calculating the average distance of the strong product directed graph $G_1 \boxtimes G_2$ by using the algorithm can be mainly divided into the following two steps:

(1) **S7 − S9:** By using the Floyd algorithm, we calculate the distances between all ordered pairs of vertices in digraphs $G_1$ and $G_2$, and store them in matrices $D_1$ and $D_2$. For $i = 1, 2$, $D_i$ is initialized to the adjacency matrix of $G_i$. And if there is a non-diagonal element with a value of zero, then let the value of the element be infinite. For each pair of vertices $m$ and $n$ of $G_i$, if there is an intermediate vertex $k(1 \leq k \leq v_i)$ such that the distance from $m$ to $k$ to $n$ is shorter than the distance of the known path, i.e. $D_i[m][n] > D_i[m][k] + D_i[k][n]$, then we update the distance from $m$ to $n$, that is $D_i[m][n] = D_i[m][k] + D_i[k][n]$.

(2) **S4 − S6:** According to the matrices $D_1$ and $D_2$, combined with the distance formula of strong product digraphs, we can calculate
$$\sigma(G_1 \boxtimes G_2) = \sum_{x_1 \in V_1} \sum_{y_1 \in V_1} \sum_{x_2 \in V_2} \sum_{y_2 \in V_2} d_{G_1 \times G_2}(x_1 x_2, y_1 y_2)$$
$$= \sum_{x_1 \in V_1} \sum_{y_1 \in V_1} \sum_{x_2 \in V_2} \sum_{y_2 \in V_2} \max\{d_{G_1}(x_1, y_1), d_{G_2}(x_2, y_2)\}.$$

The process of calculating $\sigma(G_1 \boxtimes G_2)$ in the above equation corresponds to step (5) in **S5**, and then we calculate the average distance $\mu(G_1 \boxtimes G_2)$ of $G_1 \boxtimes G_2$.

# 4 Conclusion

The strong product method is an efficient and important method for constructing a large digraph through some small digraphs. The topological structure of the large digraph constructed by the strong product is closely related to that of the factor digraphs. Considering the influence of distances of factor digraphs on the strong product digraph, we give a distance formula of strong product digraphs. Based on the formula, we study the average distance of strong product digraphs, and give an algorithm for calculating the average distance of strong product digraphs. There are still many undetermined graph parameters and structural characteristics of strong product digraphs, such as radius, Wiener index, Hamiltonity, etc. The distance formula of strong product digraphs provides a theoretical basis for future research.